# The accretion luminosity of a massive black hole in an elliptical galaxy

A.C. Fabian and M.J. Rees
Institute of Astronomy, Madingley Road, Cambridge CB3 0HA

**ABSTRACT**
Most large elliptical galaxies should now host a massive black hole leftover from an earlier quasar phase. Such galaxies also have an extensive hot gaseous halo. Here we consider why the nuclei of elliptical galaxies are not luminous sources due to the accretion of the hot gas by the central black hole, yet have weak radio sources. In particular we suggest that accretion from the hot medium has low radiative efficiency and forms a hot ion torus surrounding the black hole. Synchrotron emission from the torus can account for at least part of the weak radio sources commonly seen in otherwise normal elliptical galaxies. The inner funnel of the ion torus can also begin the collimation of a narrow jet. We speculate on the difference between jets from the nuclei of spiral and elliptical galaxies and suggest that the extensive hot atmosphere seen in ellipticals is essential for maintaining the collimation and provoking the observed radio structures of radio galaxies. We also note that low radiative efficiency, and not necessarily a large change in accretion rate, may be responsible for the demise of quasars and may contribute to the hard X-ray Background. Hot ion tori may be common in galactic nuclei.

**Key words:**  galaxies – active, jets, nuclei; accretion, accretion discs

## 1 INTRODUCTION

The host galaxies of luminous radio-loud objects – radio-loud quasars and radio galaxies – appear to be giant elliptical galaxies which, if the luminosity was due to accretion, should now have built up massive black holes. Giant elliptical galaxies at the present epoch also contain an extensive hot halo (Forman et al 1979; see Fabbiano 1989 for a review) which will surround the black hole and provide a source of matter for continuing accretion. This leads to a puzzle (Fabian & Canizares 1988), since the accretion rate onto a massive black hole in a nearby elliptical galaxy (using Bondi's 1952 formula and the X-ray inferred core gas densities) would make them much more luminous than they appear if the radiative efficiency were 10 per cent. NGC 4472 for example, which is optically the most luminous elliptical galaxy in the Virgo cluster, would then harbour a black hole of at most a few times $10^7 \, M_\odot$, whereas estimates from quasar counts require black hole masses above $10^8 - 10^9 \, M_\odot$ (Fabian & Canizares 1988).

The problem is more general than just one or two special objects since the straightforward application of the Bondi formula indicates that *most* giant elliptical galaxies should now host active nuclei emitting more than $10^{43}$ erg s$^{-1}$ (see also Fabbiano, Gioia & Trinchieri 1989). Available data indicate that the X-ray luminosity of elliptical galaxies rarely exceeds $10^{42}$ erg s$^{-1}$, of which only a small fraction can be from a nucleus since the bulk of the observed emission is extended. Nevertheless, low-luminosity radio activity is common (Sadler et al 1989; Wrobel & Heeschen 1991), indicating that there is some form of active central engine in the cores of these galaxies.

Various possible solutions to the problem were discussed by Fabian & Canizares (1988) including emission in some unobservable band or mostly as kinetic energy (jets), outflows or relativistic gas surrounding the hole preventing accretion, angular momentum preventing the gas from reaching the hole, time dependence and low efficiency. Here we re-examine the possibility that the radiative efficiency is low, in the light of the recent revival of advection solutions (Begelman 1978; Begelman & Meier 1982; Abramowicz et al 1988) for accretion disks by Narayan and co-workers (see Narayan & Yi 1995 and references therein). When $\dot{M}$ is low, the radiative efficiency in the (low density) accreting material may be poor. The energy transported by viscous friction then gets advected inwards. Early work on the resulting hot ion tori by Shapiro, Lightman & Eardley (1976) and Rees et al (1982) raised the question of why there is no mechanism (plasma instability say) coupling the electrons and ions more efficiently than 2-body interactions (see e.g. Begelman & Chiueh 1988). The apparently successful application of advection-dominated disks to the Galactic Centre (Rees 1982; Narayan, Yi & Mahadevan 1995) and elsewhere (see Narayan 1995) suggests that there may be no such mechanism and that advection solutions are of wide importance.

Accretion of hot gas in an elliptical galaxy may create the ideal circumstances for such disks to operate. At least some of the apparent demise of the luminosity of radio-loud objects may then be due to a change in accretion efficiency. Finally, we speculate on the influence that the hot halo in elliptical galaxies may have on the formation and propagation of radio jets.

## 2 ADVECTION-DOMINATED TORI IN ELLIPTICAL GALAXIES

Narayan & Yi (1995) show that advection-dominated solutions occur for accretion rates below $\dot M_{\rm crit} \approx \alpha^2 \dot M_{\rm Edd}$, where $\alpha$ is the standard Shakura-Sunyaev viscosity parameter and $\dot M_{\rm Edd}$ is the accretion rate corresponding to the Eddington limit. If $\alpha = 0.1$ then advection can dominate the flow if the accretion proceeds at less than one per cent of the Eddington rate. At low accretion rates the ions cannot cool and flow inward carrying an increasing amount of thermal energy. Pressure forces then maintain a thick torus at small radii and rotation is non-Keplerian. Of crucial importance is the inner boundary which must be such that matter with high internal energy can fall inward without releasing further energy. This precludes a central object with a hard surface. Moreover it requires a non-Newtonian gravitational field with the property that the energy and angular momentum of circular objects increases with decreasing radius $r$ close to the centre. For a Schwarzschild black hole, circular orbits have zero binding energy at $4r_{\rm g}$ (where $r_{\rm g}$ is the gravitational radius of the hole). If the hot ion torus has its inner boundary just outside $4r_{\rm g}$, matter with very low binding energy can be accreted into the black hole.

In Seyfert galaxies (which are mostly spirals), accretion is inferred to occur via a thin disk and with high radiative efficiency. In the cores of giant elliptical galaxies where the black holes are more massive and the accretion is proceeding from the hot interstellar medium (which should have a relatively low angular momentum) then the resulting quasi-Bondi flow which is already hot may immediately go to the hot, advection-dominated solution. The disks in Seyfert galaxies which radiate away the liberated gravitational energy locally can still exist at low mass flow rates; an important issue is why a system finds which solution if both are possible (cf. Narayan & Yi 1995b). The gas density may play a crucial role such that if it starts high and the temperature low it may remain there (Seyferts) and vice versa (nuclei of ellipticals).

### 2.1 Angular Momentum

A hole exactly at rest in a quiescent medium would accrete material that started off at progressively larger distances, so the specific angular momentum of gas crossing the Bondi accretion radius $r_{\rm B}$ (of order $GM/c_{\rm s}^2$, where $c_{\rm s}$ is the external sound speed) would increase steadily with time. We assume that angular momentum does not dominate the flow at that radius because either angular momentum in the hot gas is transported outward in the halo by turbulence (Nulsen, Stewart & Fabian 1984) or the hole is moving with respect to the gas. If the gas is moving subsonically past the hole with a Mach number $\mathcal{M}$, then material is accreted from a cylinder whose radius exceeds $r_{\rm B}$ by $\mathcal{M}^{-1/2}$. We suppose that this specific angular momentum is a fraction $x(<1)$ of what would be required for a Keplerian orbit at $r_{\rm B}$. The inflow across $r_{\rm B}$ would then be spherically-symmetrical. The effects of angular momentum would, however, become important within a radius $r_{\rm T} = x^2 r_{\rm B}$ Angular momentum would preclude radial accretion directly into the hole unless $x$ were smaller than $(c_{\rm s}/c) \approx (kT/m_p c^2)^{1/2}$. Within $r_{\rm T}$ (which we from now on assume is indeed larger than the gravitational radius of the hole), the flow would develop into a disc or torus (discussed below).

### 2.2 Magnetic Fields

The magnetic field strength may be close to equipartition, even outside $r_{\rm B}$, if the gas has participated in a cooling flow (Soker & Sarazin 1990). Even when angular momentum is unimportant, the symmetry and rate of the inflow would be modified by the field, especially if it is uniform on scales as large as $r_{\rm B}$ (a field could even stifle the accretion flow). An initially weak field would be amplified within $r_{\rm B}$, even by spherical inflow (Shapiro 1973, Meszaros 1975) – the combination of transverse compression and radial compression can enhance the magnetic stresses in proportion to $r^{-4}$, whereas gas pressure goes no more steeply than $r^{-5/2}$. So a weak field may have built up to equipartition even before the flow gets in to $r_{\rm T}$. However, when rotation is important, the shear will be even more effective in amplifying the field (moreover, even if there is already an equipartition field in the radial inflow, the field strength will be higher within $r_{\rm T}$, because the density is higher at radii when the inflow is slower than free-fall).

### 2.3 Radiation from the torus

If viscosity were very low, the material within $r_{\rm T}$ could settle into a dense thin disc. However, the viscosity associated with a near-equipartition magnetic field would allow a self consistent and stable inflow via a thick torus. Unless $x^2 r_{\rm B}$ is very small, there would be a range of radii where the flow is approximately self-similar (cf Begelman and Meier 1982, Narayan & Yi 1995a). There may be outflow at high latitudes and inflow near the equatorial plane. Even though the radial flow is slower than free-fall, it may still, for low-accretion rates, be faster than the cooling time. Except very near the rotation axis, (where, as discussed in Section 2.5, a fast jet may flow out along a relatively empty channnel) the radial structure in these 'advection solutions' resembles the spherical inflow for the same accretion rate, except that the densities and flow timescales are raised by a factor of order $\alpha^{-1}$.)

Physical conditions in an advection-dominated flow have recently been considered by Narayan et al (1995) on the basis of two simplifying assumptions:

(a) electrons and ions are coupled only by two-body (Coulomb) effects;
and
(b) the processes are 'thermal' in the sense that the energy dissipated is steady and sets up a roughly Maxwellian distribution for each species.

The ion and electron temperatures then become unequal in the inner regions (cf. Shapiro et al 1976); electrons radiate by bremsstrahlung and by the synchrotron process, but this radiation cannot drain all the energy from the ions because Coulomb coupling is weak, so the ions remain at the virial temperature, and can maintain a thick torus even near the hole where this is tens of MeV (cf. Rees et al 1982).

Bremsstrahlung yields a low luminosity varying, for a given $\alpha$, roughly as the square of the accretion rate. The synchrotron emission by thermal electrons also varies the same way (since the field strength would be proportional to $\dot{M}$) except so far as self-absorption occurs. Assumption (b) above may, however, be unrealistically simple – in a flow with a shearing near-equipartition field, dissipation may occur in transient, localised regions (behind shocks, or the sites of reconnection) where small numbers of electrons achieve ultrarelativistic energies; these electrons would then cool rapidly, emitting synchrotron radiation extending to higher frequencies than would arise from a purely thermal distribution. The efficiency of this process would be higher (though would still be low if the energy goes mainly into ions).

Because of uncertainty about assumption (b), we cannot confidently predict the radio spectrum or intensity. However, we expect the torus to be a detectable compact radio source. Its emission would be isotropic, except for the small doppler asymmetry due to the orbital motion (which is sub-keplerian in the self-similar part of a thick torus). Compact components that are not jets may be due to this inflow. They would be oblate. The spectrum (and particularly its extension towards the infrared and still higher frequencies) depends on how efficiently a tail of non-thermal particles can be accelerated. Inverse Compton processes would be unimportant, since the synchrotron radiation would itself have a much lower energy density than the magnetic field.

In this context it is interesting to note that Sadler et al (1989) and Wrobel & Heeschen (1991) find that most elliptical galaxies have a radio-emitting nucleus. Slee et al (1994) show that the size of most of these radio cores is on the parsec scale or less and so are powered by some engine similar to that in more luminous radio galaxies. In the picture presented here, at least some of the emission could be due to the torus.

Narayan et al (1995) have successfully applied an advection-dominated solution to accretion onto a black hole in our Galactic Centre. The issue there is similar to the one addressed here: the existence of the black hole is plausible ($M \sim 10^6$ M$_\odot$ in the Galactic Centre from spectral measurements of the dynamics of stars and gas in the vicinity; Genzel, Hollenbach & Townes 1994) yet there is a reasonably large supply of gas or detectable hard X-ray emission. The resulting radio–to–hard-X-ray spectrum they obtain is roughly flat in $\nu L_\nu$ and the radiative efficiency less than 0.1 per cent.

In a realistic inflow, there may well be a multi-phase structure (and even a thin-disc component, or a population of cool magnetically-confined clouds with a substantial covering factor). Thermal X-rays would also indicate a thermal component. An interesting example may exist in the case of M87, which contains a $3 \times 10^9$ M$_\odot$ black hole (Ford et al 1995; Harms et al 1995). If the density and temperature of its local interstellar medium are assumed (Schreier et al 1982; Fabian & Canizares 1988) then the nucleus X-ray source is several orders of magnitude fainter than expected from the Bondi solution. The flow there may thus be partially advection-dominated and the well-known jet associated with the hot-ion torus (see Section 2.5).

### 2.4 The demise of quasars and the X-ray Background

If we are correct in surmising that most large elliptical galaxies contain a massive black hole which is accreting gas with low radiative efficiency then the demise of radio-loud quasars must in part be due to a change in efficiency. This may also occur for radio-quiet quasars which are usually found in spiral galaxies. The consequences of this are considerable, since there may be considerable 'unseen' accretion proceeding around us and the masses of massive black holes may be considerably higher than previously estimated (say on the basis of the arguments used by Soltan 1982). Seyfert galaxies may just be those objects where the accretion takes place via the high radiative-efficiency solution.

The Compton-scattered radiation from hot tori is characteristically hard, since the system is photon-starved. It resembles the situation invoked by Zdziarski (1988) in his Comptonization model for the X-ray Background. Indeed if most galaxies have such a hard accretion source with a mean luminosity of about $10^{40}$ erg s$^{-1}$ then the problem of the origin of the (hard) X-ray Background (see Fabian & Barcons 1992 and references therein) could be solved. The example of our own Galaxy shows that not all galaxies have a hard nucleus at that luminosity, but present data cannot rule out a majority of galaxies with such a nucleus, perhaps if evolution is included.

### 2.5 Jets

There are two other important effects of the thick torus: (i) as already mentioned briefly, the inner boundary condition must adjust itself so that material can flow into the hole from orbits whose specific binding energy is the same (small) fraction of $c^2$ as the amount of rest-mass energy radiated during the inflow – this will be much less than the 6 – 42 per cent characteristic of thin-disc accretion, and requires the matter to be swallowed from orbits closer in (and with more angular momentum) than the most tightly bound stable orbit, (ii) near the rotation axis, there is a (roughly paraboloidal) region where axisymmetric flow cannot be maintained by any balance between gravity and pressure forces, because gas in that region would have so little angular momentum that it could fall directly into the hole (Fishbone & Moncrief 1976; Fishbone 1977). If inflow occurs in this region, the gas density would be no higher than in the radial accretion case (i.e. lower by $\alpha$ than away from the axis) – this 'vortex' region therefore forms a natural channel which could readily be evacuated by an outflowing jet.

The jet could be energised by loops of field anchored in the torus itself. However, energy could be extracted, perhaps more efficiently, from the hole itself via the Blandford-Znajek mechanism (cf. Rees et al 1982). This process depends on the spin of the hole, and on the strength and structure of the applied field. This mechanism for jet production, which channels energy directly into Poynting flux and ultra-relativistic electron-positron pairs, could be relatively more important in the low accretion rate sources (where dissipation within the torus itself is less efficient)

A relativistic jet might therefore be another feature of these accretion flows – the energy discharged in this way depends, however, not just on the nature of the black hole, but on whether the spin of the hole is aligned with that of the surrounding torus. Note that Compton drag on a jet from a low-radiative-efficiency torus would be less, and thus the bulk Lorentz factor achieved by the jet could be higher, than is the case from a standard, high-efficiency, thin accretion disk. (We note that the situation discussed here may be also relevant to the beaming model for BL Lac objects; a lack of strong optical emission lines could in part be due to a lack of strong ionizing radiation from the accretion flow.)

## 3 A CONJECTURE ON THE HOT INTERSTELLAR MEDIUM AND THE COLLIMATION OF JETS

A long-standing problem is what connects the existence of jets to the type of host galaxy. Radio loud objects occur predominantly in early-type galaxies, which also have hot haloes of gas. This can be the catalyst for jets either through the formation of a hot-ion torus, as just discussed, or through the effects of its extended high pressure. Close to the centre of these galaxies the pressure due to the weight of overlying gas can exceed $nT = 10^7$ cm$^{-3}$ K. This is much higher than found in late type galaxies, which tend not to show highly collimated jets, although they often contain active nuclei (Seyfert 1 galaxies for example).

We suggest that *the surrounding gas pressure is important to the production and collimation of narrow jets*. It is possible that all active nuclei produce outflows but only in the high pressure environment of early-type galaxies are they collimated to produce narrow radio jets. The weak radio sources in spiral (and Seyfert) galaxies are different to those in elliptical galaxies (Sadler et al 1995). Of course, an extensive high pressure environment also provides a good working surface for the jets, thus improving their detectability.

The only evidence for such high interstellar pressures in spiral galaxies is found in our Galaxy and is circumstantial. Spergel & Blitz (1992) derive the pressure principally from the presence of hot X-ray emitting gas in the Galactic Bulge (Yamauchi et al 1991). Such gas cannot be bound and so might either be flowing rapidly out of the Bulge or have a low filling-factor (being say in magnetically bound filaments). The mass loss rate implied by hot space-filling matter with a scale height of 1 kpc, as Spergel & Blitz suggest, is about 15 M$_\odot$ yr$^{-1}$. This clearly cannot be a long term property of the Bulge, but must be a very brief transient phase. This also applies to any general extensive high-pressure gas in the bulges of spiral galaxies.

The Bondi-accretion radius for a $10^9 M_9$ M$_\odot$ black hole in an elliptical galaxy containing a hot ($\sim 10^7$ K) atmosphere is about $10^{20}$ cm which greatly exceeds the collimation length for a radio jet (see e.g. Biretta 1993). Consequently in the picture we present, the collimation takes place, or is aided by, the quasi-spherical accretion flow itself. The ram pressure (and if cooling is inefficient the thermal pressure) of that flow varies as $r^{-5/2}$, so rapidly rising to very high values within the accretion radius from the outer value of $\sim 10^7$ cm$^{-3}$ K to $5 \times 10^{10} M_9^{5/2}$ cm$^{-3}$ K at one pc. Beyond the accretion radius the pressure in the cooling hot atmosphere varies roughly as $r^{-3/2}$. In the most luminous radio galaxies (such as Cygnus A), the surrounding atmosphere is part of an intracluster medium which both enhances the central pressure above what is indicated above and produces a much more extensive pressure gradient.

## 4 ACKNOWLEDGEMENTS

We thank Ramesh Narayan, the referee, for helpful comments and the Royal Society for support.